\documentclass{nature}
\usepackage{graphicx}
\usepackage{dcolumn}
\usepackage{bm}
\usepackage{ifpdf}

\title{High efficiency coherent optical memory with warm rubidium vapour.}
\author{M. Hosseini, B.M. Sparkes, G. Campbell, P.K. Lam and B.C. Buchler}

\begin{document}
\maketitle

\begin{affiliations}
 \item ARC Centre of Excellence for Quantum-Atom Optics,  Department of Quantum Science, The Australian National University, Canberra, Australia. Correspondence and requests for materials should be addressed to M.H. (email: m.hosseini@anu.edu.au).
\end{affiliations}

\begin{abstract} 
By harnessing aspects of quantum mechanics, communication and information processing could be radically transformed. Promising forms of quantum information technology include optical quantum cryptographic systems and computing using photons for quantum logic operations. As with current information processing systems, some form of memory will be required.  Quantum repeaters, which are required for long distance quantum key distribution, require quantum optical memory as do deterministic logic gates for optical quantum computing. Here we present results from a coherent optical memory based on warm rubidium vapour and show 87\% efficient recall of light pulses, the highest efficiency measured to date for any coherent optical memory suitable for quantum information applications.  We also show storage recall of up to 20 pulses from our system. These results show that simple warm atomic vapour systems have clear potential as a platform for quantum memory.

\end{abstract}

  \maketitle
 An ideal quantum memory must be capable of coherently storing multiple quantum states of light for on-demand recall with memory fidelity beyond the classical limit. Arriving at this goal is a challenge for experimentalists and extensive research efforts have been dedicated to the development of  such a quantum memory for the last decade. The motivation for this activity is the promise of revolutionary quantum information technologies.  Quantum key distribution (QKD) is already a proven technique for the secure distribution of cryptographic keys\cite{Gisin:2002p11900,Gisin:2007p11898}, but practical implementations are limited to distances on the order of 100km by the transmission losses in optical fibre or the atmosphere.  Quantum computing based on optical processes\cite{Knill:2001p10177} has been shown to work in principle\cite{Politi:2009p12052} but this technology is limited in scale by the probabilistic nature of the optical quantum gates.  A practical optical quantum memory could overcome the current limits to QKD\cite{LMDuan:2001p7084} and optical quantum computing\cite{Kok:2007p9997}.
 
 Many protocols have been proposed in order to realise such a memory device, these include electromagnetically induced transparency (EIT)\cite{Fleischhauer:2000p7451}, off-resonant Raman interactions\cite{Reim:2010p11902}, controlled reversible inhomogeneous broadening (CRIB)\cite{Moiseev:2001p4503, SAMoiseev:2003p4515,Nilsson:2005p4548}, atomic frequency combs (AFC)\cite{Afzelius:2009p11906}, and spin-polarization\cite{Julsgaard:2004p7617}. Of these techniques, the most impressive efficiencies so far attained are 43\% using EIT\cite{Phillips:2008p9536} and 35\% using AFC\cite{Amari:2010p11909}.
 
In our work we use the Gradient Echo Memory (GEM) scheme. The key to GEM is the use of an external field that induces a linear atomic frequency gradient in the direction of propagation.  The frequency gradient means that the spectral components of the input light are encoded linearly along the length of the cell.  Modelling has recently shown how this frequency encoding nature of GEM can be used to manipulate stored information allowing spectral compression, frequency splitting and fine dispersion control of pulses while they are stored in the memory\cite{Buchler:2010p11952}. Recall is achieved without $\pi$-pulses simply by reversing the field gradient (see Fig.~\ref{setup}a).  This technique has been implemented in two-level praseodymium ions where recall efficiencies of 69\% have been shown\cite{Hedges:2010p11910}. In this solid state system, the frequency gradient is applied using an electric field to induce a Stark shift. 
Furthermore using AFC it has been demonstrated that optical information can be mapped into ground states of praseodymium atoms doped into crystal\cite{Afzelius:2010PRL104}. The GEM scheme has also been adapted to three-level $\Lambda$ structured atoms (Fig.~\ref{setup}b). This protocol makes it possible to reorder a stored pulse sequence allowing recall of any individual pulse at any chosen time\cite{Hosseini:2009p8466}.  This approach could work as an optical random-access memory for quantum information encoded in time-bin qubits.   The $\Lambda-$GEM scheme uses long-lived atomic ground states for storage allowing a substantial increase of memory lifetimes compared to two-level GEM\cite{Hetet:2008p5840, Hetet:2008p7696}. Using ground state coherences also enables a wider range of atomic systems, such as alkali atomic ensembles, which are easy to address with diode laser systems and can be contained in simple off-the-shelf vapour cells. Furthermore, it has recently been  shown that vapour cells can be prepared with a single compound alkene based coating in order to achieve spin relaxation times of the order of few seconds \cite{Balabas:3p12244}. \\
Here we present results of  a $\Lambda-$GEM  experiment where memory efficiencies greater than 80\% have been coherently measured from a warm rubidium vapour. We also report on the storage and recall of 20 pulses. The current results show great potential of this memory scheme for storage of quantum states of light.

\section*{Results} 

 The experimental setup is shown in Fig.~\ref{setup}c. To prepare the signal beam, a fibre coupled phase-modulator was used to produce sidebands at the ground-state hyperfine splitting of $^{87}$Rb (6.8~GHz). The beam was  sent to a cavity (finesse=100) resonant with the +6.8~GHz sideband to filter out the carrier and the -6.8~GHz sideband. Both the coupling and signal beams were intensity controlled using acousto-optic modulators (AOMs).  Signal pulses were prepared with a peak power of $ < 2\ \mathrm{\mu W}$.  The signal pulses and control beam were then combined with the same linear polarisation using another cavity (also finesse=100) resonant with the signal field.  The control and signal fields were converted to circular polarisation and sent into the gas cell. After the cell, the signal field was coupled to a single mode fibre and sent to a heterodyne detection system. The coupling of the control field to this fibre was weak due to its larger mode diameter (for more details see Methods section).

\begin{figure}[b!]
  \centerline{\includegraphics[width=\columnwidth]{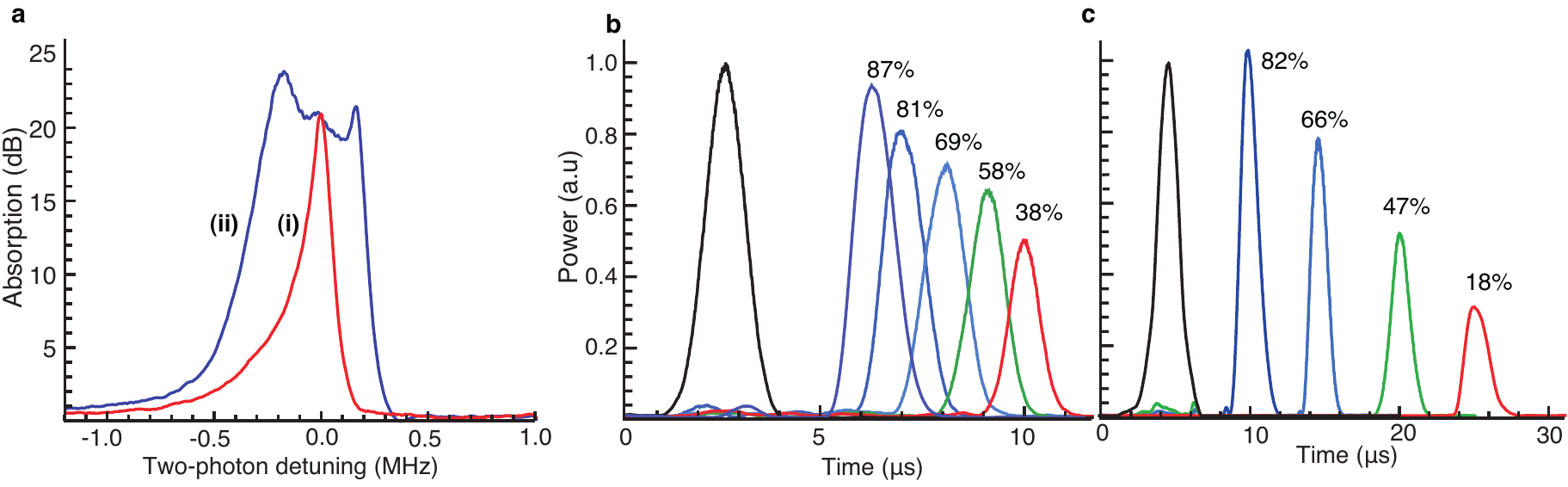}}
  \caption{\textbf{Raman Absorption line and input-echo pulses.} \textbf{a} (i) The Raman absorption line before broadening. (ii) The Raman absorption line after application of the magnetic field gradient. This was observed with a single frequency cw signal beam, while the frequency of the control beam was scanned.  \textbf{b} Storage and recall data with an input pulse duration of 2$\mu$s. \textbf{c} Storage and recall data with an input pulse duration of 3 $\mu$s and the control field is turned off during the storage time to reduce the decay rate of the storage. For both \textbf{b} and \textbf{c} the far off-resonant transmitted input pulse, which is used to normalise our recall efficiency, is shown in black. The control field power was 370~mW.}
  \label{Raman_echo}
  \end{figure}

Figure \ref{Raman_echo}a shows the Raman absorption line as a function of two-photon detuning (detuning from the Raman resonance)  (i)  with and (ii)  without the applied magnetic field gradient. The absorption is sensitive to alignment, which was optimised for for the broadened feature. With the applied broadening the absorption is $\sim$99\%. This limits the maximum possible recall efficiency of our memory\cite{Longdell:2008p8530} to 0.99$^2$=98\%.

\textbf{Pulse storage and recall.} The results of  storage and recall experiments are shown in  Fig.~\ref{Raman_echo}b.  The input pulse, shown in black, has a $1/e^2$ width of  2$\mu$s. We measure the power in this input pulse by recording the far off-resonance transmission through the gas cell without the control field.  We measured no noticeable absorption from the atomic ensemble under these conditions and we can use the total energy in this pulse (the area under the curve) to normalise the recall efficiency of our storage experiments.

In Fig.~\ref{Raman_echo}b after flipping the magnetic field gradient we recall the signal light with a maximum efficiency of 87\% .  The storage time in this case is 3.7$\mu s$ peak-to-peak, or exactly one pulse width between the $1/e^2$ points power levels to ensure complete separation of the input and recalled pulses.  The recall efficiency drops rapidly for longer storage times, although for these data the control field was on at all times.  In Fig.~\ref{Raman_echo}c we show results of recall experiments where the control field was switched off during the storage phase of the experiment.   In this case we find slower decay of the recalled pulse since we have now reduced the decoherence caused by the control field. The pulses in these data are slightly compressed on recall due to a higher magnetic field gradient used to recall the signal light, which is the reason for the peak recalled power exceeding the input peak power. We can achieve slightly higher efficiency using compressed pulses since the total storage time in the medium is reduced.

\begin{figure*}
     \includegraphics[width=16cm]{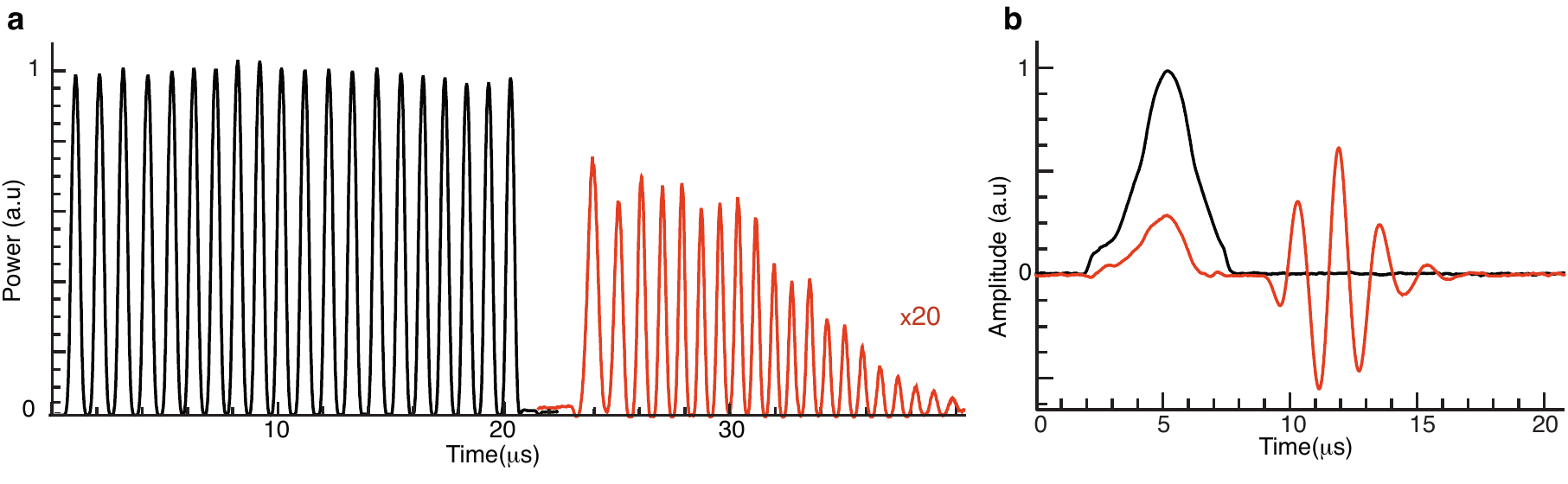}
   \caption{  \textbf{Results of 20 pulse storage and frequency shifting.} \textbf{a} Amplitude of 20 Gaussian input (black) and echo pulses (red) with total recall efficiency of 2 $\%$. \textbf{b} The input pulse (black) is stored and recalled with an introduced offset to the magnetic field gradient.  The red curve denotes the straight transmitted pulse and the recall pulse. The interference pattern shows that the pulse is coherent with the original light field and is shifted by 600kHz.}
  \label{RG-20-Fqsh}
 \end{figure*}

    In Fig.~\ref{RG-20-Fqsh}a we show the storage and recall of 20 Gaussian pulses with an overall efficiency of $2 \%$.  For this experiment the control field power was reduced from 370~mW to 64~mW to reduce the decay rate of the memory.  The lower optical depth in this case limits the efficiency of the storage and recall.  From this data we can infer a delay-bandwidth product (DBP)\cite{Tidstrom:2007p12004} of $\sim$40 for the our memory.  This is comparable to a recent demonstration of 64 pulses delayed using AFC with efficiency of 1.3$\%$\cite{Usmani:2010p11914}. Fig.~\ref{RG-20-Fqsh}b demonstrates frequency shifted recall of a pulse\cite{Buchler:2010p11952}. This was achieved by applying an offset magnetic field after the magnetic field gradient was flipped to increase the splitting of the atomic ground states.  On recall the pulse is shifted by the added splitting, which in this case is 600~kHz, as seen by the interference fringes in the heterodyne signal. The presence of interference fringes proves the coherent recall of our memory. 

\section*{Discussion}
To understand the limits of our atomic system, we now examine some theoretical aspects of the scheme. Assuming far-detuned Raman interaction, the system depicted in Fig.~\ref{setup}a can be treated as a quasi two-level atom. One can write the Maxwell-Bloch equations for the atom-light interaction scheme depicted in Fig.~\ref{setup}a in terms of atomic coherence $\sigma_{12}$ and electric field amplitude $\mathcal{E}$,
\begin{eqnarray}\label{ANALGEM}
\dot{\hat{\sigma}}_{12} &=&  -(\gamma_{12}+i\delta(z,t))\hat{\sigma}_{12}- ig\frac{\Omega_c}{\Delta}\hat{\mathcal{E}}_s  \nonumber \\
\frac{\partial}{\partial z}\hat{\mathcal{E}}_s&=& i\mathcal{N}\frac{\Omega_c}{\Delta}\hat{\sigma}_{12},
\label{eq:one}
\end{eqnarray}
where $\mathcal{N}=\frac{gN}{ c}$ is the effective linear atomic density, $g$ is the atom-light coupling, $\delta(z,t)$ is the additional splitting between the two ground states (two-photon detuning) due to the magnetic gradient, $\Omega_c$ is the control field Rabi frequency, and $\Delta$ is detuning from the excited state. The dephasing rate of $|1\rangle$ to $|2\rangle$, $\gamma_{12}$ in Eq.\ref{eq:one}, includes population shuffling due to atomic collision and other dephasing mechanisms such as the power broadening\cite{Citron:1977p11915} caused by Spontaneous Raman Scattering (SRS). 
Based on our experimental conditions, the rate of spontaneous emission due to the signal beam is found to be 2$\pi \times1$~Hz at pulse peak intensity. The scattering rate due to the control field, on the other hand, is found to be $\sim 2\pi\times 30$~kHz at $\Delta=3$~GHz and a power of 370~mW.

In the case of EIT systems, the effect of four-wave mixing (FWM) attributed to the interaction of the control field with the $F = 1 , m_F = 1$ level is proven to affect memory performance at large optical depths\cite{Phillips:2009p11917,Phillips:2008p9536}. In these experiments a double$-\Lambda$ transition appeared in the system aided by the presence of a seed Stokes  field at -6.8~GHz generated by the EOM. In our experiment the -6.8~GHz sideband                seed is rejected by two cavities that are on resonance only with the +6.8~GHz sideband, thereby eliminating any contribution of coherent FWM to our system.

\begin{figure}[h!]
  \centerline{\includegraphics[width=16cm]{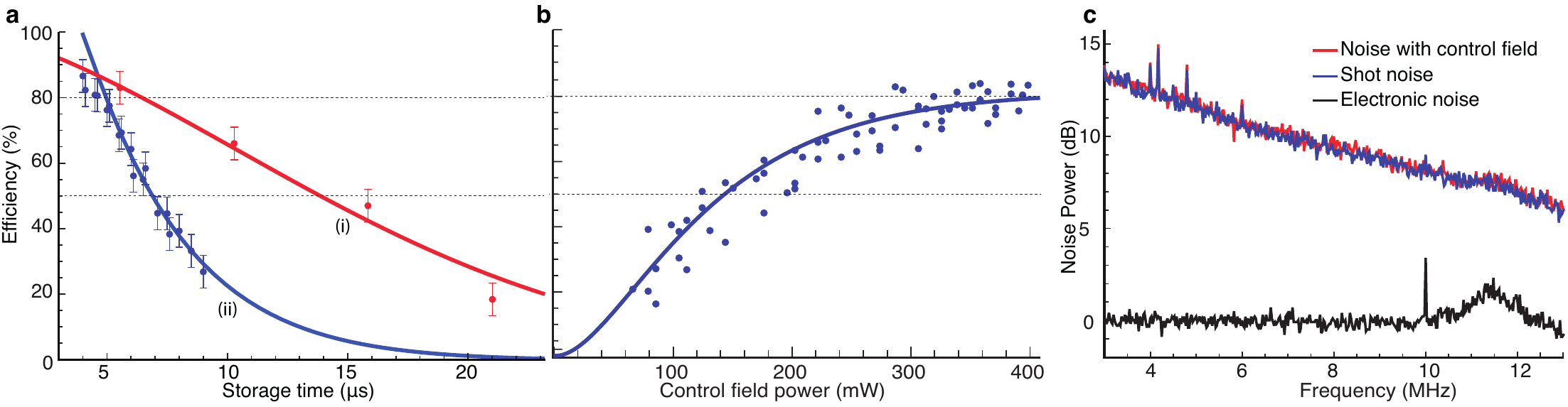}}
  \caption{\textbf{Control field effect on efficiency and noise}. \textbf{a} Echo efficiency as a function of storage time. (i, red) data taken for 3 $\mu s$ pulse while the control field with power of 380 mW was switched off during the storage time. (ii, blue) data taken for 2 $\mu s$ pulse while the control field with power of 290 mW was kept on during the storage time. Error bars indicate the detection error derived from fluctuation of the amplitudes of pulses. \textbf{b} Efficiency of photon echoes of a 3 $\mu s$ pulse as a function of control field power. The solid line is the theoretical predictions taking into account diffusion time of 22 $\mu s$, control field induced scattering and ground state decoherence rate of 2$\pi\times$3.5 kHz.  \textbf{c} Variance of the probe field mode measured using heterodyne detection. Curves represent  electronic noise (black), shot noise (blue) and noise with the control field switched on (red). Measurements were made with a Resolution Bandwidth=3 kHz, Video Bandwidth=30kHz and 5 averages. The the control beam was filtered out of this measurement using and additional gas cell containing warm Rb85.}
  \label{EffPc-decay}
  \end{figure}

To understand the contribution of the control field induced scattering we investigated memory behaviour both in the presence and absence of the control field during the storage time. The control field can be switched off during the storage time without affecting the memory protocol. We can gain some insight into our system by considering a simple model that includes an atomic diffusion time, $\tau_d$, a total ground state decoherence rate, $\tau_0$, and a maximum possible memory efficiency, $\eta_0$, limited by the optical depth.  The efficiency $\eta_m$ will then be given by
\begin{eqnarray}\label{decay}
\eta_m = \eta_0 e^{-(t/\tau_d)^2} e^{-t/\tau_0}
\end{eqnarray}
Fig.~\ref{EffPc-decay}a(i) shows the efficiency as a function of storage time when the control field is on. Taking into account the signal beam radius of 3 mm and 0.5 Torr Kr buffer gas, one can calculate the diffusion time of the atoms, defined as the time that a fraction $1/e^2$ of atoms have moved a distance greater than the radius of the signal beam,   to be  $\tau_d= 22 \mu s$. This value was fixed in our model allowing us to fit only the ground state decay time, which was determined to be $\tau_0= 4 \mu s$ corresponding to a decay rate of $2 \pi \times 40$~kHz.  This is consistent with the scattering rate of $2 \pi \times 30$~kHz calculated above, from which we conclude that our system is limited in this regime by control-beam induced scattering. 
   
   We note, however, that this may not be the case for very short storage times where the decay is not exponential and does not agree with our simple model. This effect can be attributed to the highly photonic nature of the memory for short storage times.  The pulse, in this case, has yet to be fully mapped into the atomic spin wave.  The impact of the scattering, collisional and diffusion decay terms will vary as the light is absorbed into the atomic ensemble leading to decay of the memory that differs from the model at short times. 
    
    The scattering rate can be minimised by switching off the control field during the storage. Atomic diffusion then becomes the dominant decay mechanism, hence a more Gaussian-like decay is expected.  The curve in Fig.~\ref{EffPc-decay}a(i) is the result of a convolution of a Gaussian decay function ($\tau_d=22 \mu s$) due to diffusion and an exponential decay function due to the ground state decoherence.  The fitted exponential decay time in this case is $\tau_0= 60 \mu s$, giving a decay rate of $2 \pi \times 2.6$~kHz. This is still much higher than the collision limited ground state dephasing time, but this is expected because the decay rate in this case varies as the control beam is switched on and off over the course of the experiment.\\
   We also studied the effect of the control field power on the memory efficiency for storage times of one-pulse width. As can be seen in Fig.~\ref{EffPc-decay}b, the recall efficiency saturates with increasing control field power. This is because increasing $\Omega_{c}$ can effectively provide higher optical depth as long as $\Delta\gg \Omega_{c}$. In this case the effective atom-light coupling strength can be described as $g\prime=g \Omega_c/\Delta$.  If $\Omega_{c}$ is further increased beyond this limit, the system can no longer be described by a simple quasi two-level atomic ensemble and there is no further improvement in optical depth.
   
   The presence of the strong control field could lead to noise sources in our memory due to spontaneous emission and Raman scattering into the mode of the probe beam.  To investigate these possible noise sources we measured the noise spectrum of the probe mode as shown in Fig.~\ref{EffPc-decay}c. With no control field present we observe the shot noise of our detection system (blue) which lies 10dB above the electronic noise floor (black). With the control field switched on (red), we observe no change in the noise level recorded by our heterodyne system.  If there where photons added to the mode of the probe field, then we would see added noise around 8~MHz, which is the frequency offset of the heterodyne beam from the probe frequency.  The absence of extra noise at this frequency is strong evidence that our memory is not prone to noise sources that could impact on quantum state storage.
      The data shown in Figs.~\ref{EffPc-decay} \textbf{b} and \textbf{c}, along with the coherent detection, is strong evidence that our high efficiency recall is not due to any nonlinear process or incoherent contamination of the signal mode by the strong control beam.

\begin{figure}[h!]
  \centerline{\includegraphics[width=85mm]{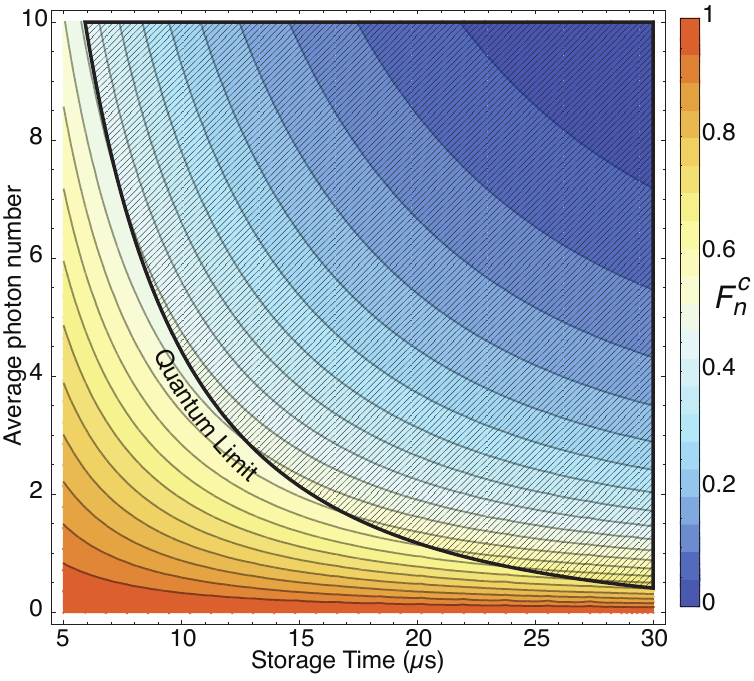}}
  \caption{\textbf{Implied quantum memory performance for coherent state storage.} The quantum limit is calculated assuming the efficiency fitted to experimental data in Fig.~\ref{EffPc-decay}a(i) and Eq.~\ref{fideq}}
  \label{fidtimes}
  \end{figure}

The best way to quantify the efficacy of a quantum memory will depend on the application.  In order to quantify how the measured efficiency of our memory would translate into a coherent state quantum memory, we can follow the model presented by He \emph{et al.}\cite{He:2009p11911} where it is shown that a linear quantum memory has fidelity ($F^c_n$)
\begin{equation}
F^c_n > \frac{1}{1+ \bar{n}(1-\sqrt{\eta_m})} \label{fideq}
\end{equation}
for coherent states with average photon number $\bar{n}$ and memory efficiency $\eta_m$. Given that our memory is linear, and assuming that no extra noise is added to the stored states, we can calculate the range of coherent amplitudes for which it can act as a quantum memory, as shown in Fig.~\ref{fidtimes}. This shows, for example that we could store coherent states up to $\bar{n}=10$ for times less than 6~$\mu s$, or states with $\bar{n}$=1 for 21~$\mu s$.

   In conclusion, we have shown light storage in warm vapour gas with more than 80$\%$ memory efficiency using $\Lambda-$GEM technique.  All measurements were made using coherent heterodyne detection. We have demonstrated a time-bandwidth product of $\sim$~40 and controlled frequency shift of the recalled pulse.  The decay rate of our memory can be controlled by minimising the use of the control beam to the extent that for longer storage times we become limited by atomic diffusion. If applied to coherent state storage, our scheme could perform as a quantum memory for states containing up to 10 photons.
   
   \section*{Methods}
\begin{figure}[b!]
  \centerline{\includegraphics[width=100 mm]{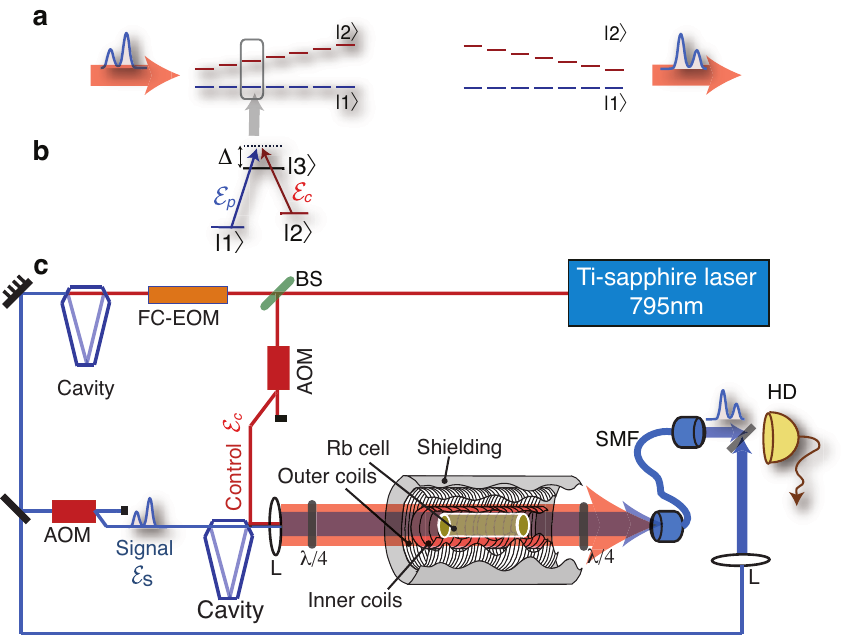}}
    \caption{\textbf{Schematic view of the experiment.} \textbf{a} Schematic view of the frequency gradient generated for an ensemble of two-level atoms. By switching the sign of the gradient from positive (left) to negative (right) a photon-echo is emitted in the forward direction. \textbf{b} Schematic view of $\Lambda$-atom structure which at large detuning is equivalent to a two-level atom based on the two ground states. \textbf{c} Experimental setup showing the control (red) and signal (blue) beams with 6.8~GHz frequency difference, collimated at radii of 6 mm and 3 mm respectively and of identical circular polarisation when they go through the cell. Heterodyne measurement is performed after the memory on the signal field. BS: Beam Splitter, HD: Heterodyne Detection, SMF: Single Mode Fibre, AOM: Acousto-Optic Modulator, FC-EOM: Fibre-Coupled Electro-Optic Modulator, $\mathcal{E}_s$ and $\mathcal{E}_c$: signal and control field amplitudes, respectively.}
  \label{setup}
  \end{figure}

\textbf{Experiment.} Our atomic ensemble consists of a cylindrical cell (length and diameter of 20 cm and 25 mm, respectively) containing $^{87}$Rb atoms mixed with 0.5 Torr of Kr buffer gas, used to increase the time of flight of  atoms inside the beams. The time of flight of atoms inside the signal and the control beam is calculated to be 45 $\mu s$ and 180 $\mu s$, respectively. We consider the off-resonant Raman interaction of a weak signal field and a strong control field, both having $\sigma^+$ polarization, addressing $|1\rangle$ to $|3\rangle$  ($S_{1/2}, F=1\rightarrow P_{1/2}, F^{\prime}=2$) and $|2\rangle$ to $|3\rangle$ ($S_{1/2}, F=2\rightarrow P_{1/2}, F^{\prime}=2$) transitions respectively, as shown in Fig.~\ref{setup}b. These fields are blue-detuned by $\Delta$=3~GHz from the excited state.

The temperature of the cell was held around 78$^{o}$C using a bifilar resistive heater wrapped around the cell. We estimate the number of atoms in the signal beam mode volume to be $\simeq4\times10^{12}$.  Considering the corresponding Clebsch-Gordan coefficients\cite{Rb87}, approximately 65$\%$ of these atoms will be located in $F = 1 , m_F = 1$ state with the rest of the population located in $F = 2 , m_F = 2$. 
To create the frequency gradient, two specially wound magnetic coils surrounding the cell were used to produce opposing linearly varying Zeeman shifts along the cell\cite{Hosseini:2009p8466}.  Switching current between these coils created a switchable atomic frequency gradient in our storage medium. To recall the pulse, one coil is switched off and one is switched on within 0.5 and 2.5 $\mu s$, respectively. The DC magnetic field is normally set to 6 G and the typical value of the gradient field is 20 mG/cm. The cell and coils are surrounded with $\mu-$metal to reduce the influence of the Earth's magnetic field.
\bibliographystyle{naturemag}
     \bibliography{bibs}
         
 \textbf{Acknowledgements}  The authors thank M.~Sellars for enlightening conversations. This work was supported by the Australian Research Council.
 
 \textbf{Competing financial interests} The authors declare no competing financial interests.

\textbf{Contributions}
Experiments, measurements and data analysis are performed by M. H. with assistance of B. M. S. and G. C. on data collection and experiment preparation.  All authors discussed the results and commented on the manuscript.  B.B. and M. H. prepared the manuscript. B.B. and P.K.L.  planned and supervized the project.

\end{document}